\documentclass[aps,pre,amsmath,amsfonts,amssymb,floatfix,superscriptaddress,nofootinbib]{revtex4}
\setcitestyle{authoryear,round}

\usepackage{graphicx}  
\usepackage{dcolumn}   
\usepackage{bm}        
\usepackage{amssymb}   
\usepackage{amsmath}   
\usepackage{color}
\usepackage{titlesec}

\newcommand\be{\begin{equation}}
\newcommand\bi{\begin{itemize}}
\newcommand\bea{\begin{eqnarray} \nonumber }
\newcommand\ee{\end{equation}}
\newcommand\ei{\end{itemize}}
\newcommand\eea{\end{eqnarray}}

\newcommand\rr{}

\def\thesection{\arabic{section}}
\def\thesubsection{\arabic{section}.\arabic{subsection}}

\titleformat{\section}
{\normalfont\large\bfseries}
{\thesection.}{0.5em}{}

\titleformat{\subsection}
{\normalfont\bfseries}
{\thesubsection.}{0.5em}{}

\begin{document}

\unitlength = 1mm
\title{A million metaorder analysis of market impact on the Bitcoin}

\author{J.~Donier}\affiliation{Laboratoire de Probabilit\'es et Mod\`eles Al\'eatoires, Universit\'e Pierre et Marie Curie (Paris 6).}
\author{J.~Bonart} \affiliation{CFM-Imperial Institute of Quantitative Finance, Department of Mathematics, Imperial College, 180 Queen's Gate, London SW7 2RH}

\date{\today}

\begin{abstract}

We present a thorough empirical analysis of market impact on the Bitcoin/USD exchange market using a complete dataset that allows us to reconstruct more than one million metaorders. We empirically confirm the ``square-root law'' for market impact, which holds on four decades in spite of the quasi-absence of statistical arbitrage and market marking strategies. We show that the square-root impact holds during the whole trajectory of a metaorder and not only for the final execution price. We also attempt to decompose the order flow into an ``informed'' and ``uninformed'' component, the latter leading to an almost complete long-term decay of impact. This study sheds light on the hypotheses and predictions of several market impact models recently proposed in the literature and promotes heterogeneous agent models as promising candidates to explain price impact on the Bitcoin market -- and, we believe, on other markets as well.

\end{abstract}

 \maketitle

\section{Introduction}

Most financial markets have undergone rapid changes in the past 10 years. The implementation of limit order books, the electronization of trade exchanges, the rise of high-frequency trading, and the introduction of algorithmic and automated executions are the most emblematic features of the new financial economy. Both supporters~\citep{Hendershott11,Hendershott14} and critics~\citep{Budish,KirilenkoKyle} of this evolution hold the view that the recent speed revolution has had a profound impact on the functioning of financial markets. While this is obviously true from a technological and microstructural point of view, it is less clear from an economists' perspective: In the end, a stock exchange is a market place where buyers and sellers meet and agree on a price; and it is not straightforward to assess the relevance of technological changes to this process.

%
It is however possible to directly compare financial markets which strongly differ in their respective degrees of technological sophistication and latency. The recent development of cryto-currencies such as the Bitcoin, traded against classical currencies on automated exchange platforms, provides us with a unique benchmark for a comparison with high-speed markets. Indeed, the Bitcoin exchange is an example of a proto-typical financial market, maintained -- at the time of the study -- in a rudimentary competitive state by high trading fees\footnote{0.6\% fees per transaction for the vast majority of the main 
exchange MtGox's users} which inhibit the development of substantial market making or arbitrage.

An important aspect of a financial market is how it absorbs the new information conveyed by a trade or a sequence of trades into the market price~\citep{kyle}. Such incorporation of new information may not be instant due to market frictions~\citep{Beja80}; nor is the submission of the trader's order, since order splitting and inventory considerations create a serial dependence of trades~\citep{FarmerLillo04,Toth15}. During the execution of a large sequential order (meta-order), spread over a certain time period, the difference between the impacted price and the initial price can be quantified by measuring the market impact of the meta-order.

In this paper we undertake a thorough empirical analysis of market impact on the Bitcoin/USD exchange market. Previous studies~\citep{almgren2005direct,moro2009market, toth2011anomalous, bershova2013non, gomes2015market, mastromatteo2014agent} found that average market impact of a meta-order of total volume $Q$ approximately follows the ``square-root law'',
\begin{equation}\label{eq:impact_emp}
I(Q) {\approx} \pm Y\sigma\left (\frac{Q}{V_D}\right )^{\delta} \;,
\end{equation}
where $I(Q)$ quantifies the average difference between the impacted price and the initial price (with the positive sign corresponding to buy orders and vice versa), $V_D$ and $\sigma$ the daily traded volume and daily volatility of the stock. The exponent $\delta$ has been consistently found to be approximately $1/2$. When the meta-order terminates so the pressure it was exerting on the price stops, the price is usually observed to revert (partially or totally) towards the initial (unimpacted) price~\citep{brokmann2014slow,gomes2015market}. 
With our high-quality dataset we shall confirm that above impact law holds for the Bitcoin/USD exchange market. Therefore, the rise of algorithmic trading may not have had as a profound effect on the functioning of markets as is often advocated. On the Bitcoin, both the response of the market to trades and the serial dependence of orders are similar to what is observed on mature liquid financial markets, in such a way that market impact can be specified by the same empirical law.

This observation is central: The precise mechanism which is behind the peculiar square-root law appears to be universal across markets which significantly differ with respect to their trade characteristics (latency, daily traded volume, volatility), microstructural parameters (tick or lot sizes) and fee structure (i.e. the high fees on the Bitcoin).
This latter point is especially important. Statistical arbitrage is not profitable on the Bitcoin on scales below the fee level of 60 bps (120 bps for a round-trip!) and price efficiency is therefore not ensured in this region. The resulting large bid-ask spread frequently exceeds the peak impact of typical meta-orders; concepts such as price efficiency and arbitrage are hence not suitable to explain market impact on the Bitcoin exchange.

Therefore, we conclude that competitive equilibria between different agents, for instance amongst market makers or between informed liquidity takers and uninformed liquidity providers~\citep{glosten}, should not be used as the fundamental starting point for the theoretical modelling of market impact on the Bitcoin. Nor should one use the martingale conditions for the market price, as the notion of ``price'' is not precisely defined under the scale of 60 bps ($\sim$spread/fees). In fact, as we shall see in the following, impact describes \emph{how trades dig into the opposite (supply/demand) side} -- before some post-trade mean-reversion occurs on this side -- rather than how the \emph{market price} itself is affected. Although these two definitions are equivalent when the spread is tight, they do not coincide in the case of the Bitcoin. The fact that very similar impact laws are found on mature financial markets suggests that this observation holds for these markets, as well.

\subsection{Comparison with related literature}

The square-root impact formula quantifies how market prices are affected by trades. This has implications for \emph{market stability}: Price impact may lead to unexpected price swings~\citep{lehalle2012sawtooth} and stock crashes~\citep{kyle2} as well as to well-known phenomena such as stock pinning~\citep{avellaneda2012mathematical}. 
Second, the control of \emph{execution costs}, i.e. market impact, is of great practical interest to financial institutions. \citet{almgren2001optimal}, for instance,  
determined optimal execution strategies based on the assumption that impact is \emph{linear} (i.e. that $\delta=1$). This is the simplest theoretical setting that excludes round-trip execution strategies (zero terminal inventory) with negative execution costs (profit making). 
Since then, a huge effort has been made to model non-linear market impact, motivated by a flurry of empirical studies. These approaches can be broadly split into three categories:
\begin{itemize}
\item Concave market impact of metaorders can be reproduced by \emph{propagator models}, in which each trade is assumed to have a transient impact which decreases according to some time-dependent kernel. Summing up the impacts of individual trades leads to the impact function of the metaorder which is then found to be concave. The possible absence of dynamical arbitrage in these settings~\citep{gatheral2010no} allows one to solve convex optimization problems for optimal execution and find optimal liquidation strategies~\citep{alfonsi2013capacitary} within this framework. While these models yield fairly realistic results and are analytically tractable, they are however purely phenomenological and do not provide a mechanism to explain impact.
\item \emph{Equilibrium models} assume a competitive equilibrium between liquidity providers and takers. They can be regarded as an extension of the original argument in~\citet{glosten} from a single market order to a sequence of trades. Equilibrium models typically use two constraints to fix the bid and the ask during the metaorder execution, usually using the martingale condition and a subtle fair pricing argument~\citep{farmer2013efficiency, donier2012market}, which states that the transient impact of a metaorder anticipates its permanent impact on the price so that neither the informed trader nor the market maker should {\rr expect profits for any} metaorder (on average). {\rr In the same vein, \citet{jaisson2015market} shows how non-linearities can emerge from anticipations when the order flow is correlated by a Hawkes process, even though impact is linear \textit{ex post}.} These arguments, together with strong correlations in the order flow finally leads to an -- asymptotically -- concave metaorder impact. {\rr However, as such mechanisms are rather unlikely to be in force on the Bitcoin market, they seem not appropriate to explain the seemingly universal shape of impact}.
\item The third class of models, initiated in a different context in~\citet{Bak:1997}, are \emph{statistical models} of supply and demand~\citep{toth2011anomalous,mastromatteo2014agent,mastromatteo2014anomalous,DonierBonart14}, that may also be seen as \emph{heterogeneous agents models}~\citep{donier2015walras}. {\rr The dynamics they give to the supply and demand that underlie the order book is such that both generically increase quadratically with distance from the mid-price}, in turn leading to an exact and universal square-root impact at all scales.
\end{itemize}

Empirical studies have been mostly concerned with the \emph{peak impact} of meta-orders, i.e. the impact measured between the extremal points of the execution path. Further studies have shown that impact is to a large part transient: after execution the price falls from its peak to some intermediate level, that some studies~\citep{moro2009market, bershova2013non, gomes2015market} argue to be close to $2/3$ of the peak impact, in agreement with equilibrium models~\citep{farmer2013efficiency}. In~\citet{brokmann2014slow} the authors find that this high permanent level may be due to the correlation between the trader's execution decision and the residual order flow\footnote{{\rr After a buy (resp. sell) metaorder, the order flow is on average biased towards buy orders (resp. sell orders). This correlation with future order flow artificially inflates its measured impact.}} and to the price signal that triggered the decision to trade: By taking into account these effects they argue that the ``bare'' permanent impact (or mechanical permanent impact) is much lower and possibly even zero. {\rr Finally, \citet{gomes2015market} conduct a separate impact study of informed trades and cash-flow (uninformed) trades, to find that the latter have no permanent effect on the price (although the transient impact are similar for both types of trades).} Hitherto, these results are not well accounted for nor understood.

Amongst the empirical studies of market impact in the past literature, the vast majority has relied on partial datasets. Usually, market participants have only 
access to their own proprietary data \citep{toth2011anomalous, bershova2013non, brokmann2014slow} which leads to an unavoidable conditioning to their trading strategies (even though in some cases many different strategies are collated together so part of the conditioning may average out). Two notable exceptions do however exist: 
In~\citet{moro2009market} hidden metaorders are directly inferred from brokerage codes, while \citet{Farmer:new} have unprecedented access to the start times, end times and volumes of a huge amount of metaorders stemming from Ancerno's clients\footnote{This study reports a log impact as a better overall fit to metaorder impact, {\rr although they actually mention some caveats due to the low level of details of their dataset}.}.

In this paper, we use a dataset which allows on the contrary to identify \emph{each} trade with a \emph{unique} trader, thereby leading to a \emph{complete picture of the market}. Our dataset is large enough to study market impact as a function of volume, participation rate and even as a function of the behaviour of the residual market. This allows us to retrieve pseudo-random metaorders, i.e. metaorders that are uncorrelated from the residual order flow, either because they do not convey any information or because the information that triggered them is not shared by the residual market.

\subsection{Brief outline}

After a presentation of the dataset (Sec.~\ref{sec:data}), we introduce the main definitions and methodology (Sec.~\ref{sec:preliminary}) with a focus on the methods we use to identify distinct meta-orders. The price impact of metaorders is introduced in Sec.~\ref{sec:impact}. The first major result of our study is that, despite its prototypical micro-structure, the square root law holds on the Bitcoin. Moreover, in Sec.~\ref{sec:further} we show that not only the peak impact of metaorders, but rather the \emph{whole impact trajectory} follows a square-root. We show unprecedented pictures of the mid-price, bid and ask
 trajectories during and after impact and discuss permanent and transient market impact as a function of the execution speed and its dependence on the residual order flow. 

We are able to differentiate the mata-orders with respect to their correlation with the residual order flow. The permanent impact is identified as the information content of meta-orders: Pseudo-random metaorders (i.e. metaorders uncorrelated with the residual order flow) have little permanent impact, in agreement with previous studies~\citet{brokmann2014slow,gomes2015market}.

Finally, in Sec.~\ref{sec:sum} we summarize our main findings and in the last section we conclude and discuss 
the implications of our empirical findings for the most common impact models so far proposed in the literature.

\section{Bitcoin market at a glance and data}\label{sec:data}

\subsection{Bitcoin : a prototypical market}

{\rr Bitcoin is a crypto-currency introduced by an anonymous programmer in 2008 \citep{nakamoto2008bitcoin}, designed to allow exchanges of money without the need of a central authority (e.g. state or bank) to enforce trust. The money is issued progressively through a process called \emph{mining} -- by analogy to gold mining -- to the people who give their computing power to help build the transactions ledger (the \emph{blockchain}). As a currency, it has some intrinsic value (even though still quite imprecise) and can be exchanged against usual currencies on organized markets.} The Bitcoin market started in 2008 and literally exploded in 2013 with a peak market capitalization above $10$B dollars, a daily number of transactions over $100$k and a daily traded volume above $100$M dollars at the end of 2013. 

{\rr Similarly to financial markets, trading takes place on an order book, where Bitcoins can be exchanged against other currencies. Notably, the role of the order book is quite important since a large fraction of the liquidity is not hidden, but actually posted in the order book. A more quantitative analysis indeed shows that typically $30-40\%$ of the volume traded during the day is already present in the order book in the morning. This is to be compared with a ratio below $1\%$ on more traditional financial markets, say stocks \citep{wyart2008relation}. The tick size, i.e. the minimal price increment between two consecutive prices at which it can be exchanged, is of USD $10^{-5}$ on MtGox so that the order book can be considered as a continuous price grid.}

Bitcoin market microstructure is quite unique for several more fundamental reasons. First, because of the very high level of fees (compared to other markets) of $0.6\%$ per transaction on MtGox, resulting in an average spread of $\approx 0.6\%$. This, to a large extent, hobbles high frequency arbitrage/market making strategies on such a market, {\rr who only take part in a few percent of the transactions: the Bitcoin market is essentially a market between end users}. Second, all relevant information is concentrated on one asset and one exchange (MtGox) on the time span considered, with very little notion of a ``fundamental price''. This is a very unique example of a single-asset economy, with little correlations with any other asset on the planet (for the time being).

\subsection{Data}

This study was realized using a database of all 13M trades trades that happened on MtGox Bitcoin-USD exchange between August 2011 and November 2013, 
in which traders are uniquely identified\footnote{The data we use was provided to us by an anonymous source, and is available upon request 
under certain conditions. The public part of this data file has 
been checked to match otherwise known price and volume data at www.bitcoincharts.com, and the private part to match all private trading data that are to the knowledge of the authors. Apart from these consistency checks, no insurance is given that 
all data is accurate. However the high precision of our findings suggests that they are.}.
Thus, we have a complete description of the market as a whole, in contrast to most hedge funds' proprietary data (reflecting their trading decision) which does not allow 
for a global analysis of trading decisions and market impact. As a consequence, we do not face the problem of the potentially strong conditioning of impact paths on the 
particular strategy of the agent. We rather have an overall insight on how agent's decisions entangle. 
Since there is very little brokerage intermediation in this market, we have direct knowledge of the actual initiators of each trade (only as an anonymized numerical ID code). One can thus assume that with very good approximation 
\textit{all metaorders can be fully identified}. This is a most valuable property for studying impact, since it allows for the 
explanatory variable -- the metaorder -- to be fully characterized.
This data quality is probably very difficult to obtain on other financial markets, because of brokerage intermediation, multiplicity of venues, and multiplicity of correlated instruments on which a trade can be executed. In addition, our dataset is the largest so far with such precision, and with full knowledge of the trajectories for the 1M identified metaorders.

\section{Definitions and methodology}\label{sec:preliminary}

\subsection{Definitions}\label{sec:definitions}

When a trader (she) wishes to bring an excess of supply or demand $Q$ on the market, she is confronted to the question of how to achieve it in a reasonable way. If the volume she wants to sell or buy is small, then she will probably do it all at once if enough liquidity 
is available on the order book. 
However, if she wants to invest or sell back a quantity that exceeds the available offer or demand (think of large speculators or professional service providers), an instantaneous execution might destabilize the market and incur her larger costs than planned: 
she therefore has to split her large order into chunks, so that the imbalance can be slowly digested by the market~\citep{bouchaud2009markets}. We refer to this \textit{total} quantity $Q$ as a \textit{metaorder}, denoting $T$ its duration and 
$\mu := Q/T$ its execution speed, and study the quantities

\bi
\item ${\cal I}_{\rm path}(r, Q, \mu)$, defined as the impact on the price of the first $r\%$ of a metaorder of size $Q$ and execution speed $\mu$. By definition $r \in [0,1]$ is the part of the volume already executed, and 
${\cal I}_{\rm path}(0, Q, \mu) = 0$ is the initial price (gauged to zero);
\item ${\cal I}(Q, \mu) := {\cal I}_{\rm path}(1, Q, \mu)$ is the price at the end of the metaorder, referred to as the \textit{peak impact} of the metaorder;
\item ${\cal I}_{\rm exec}(Q, \mu) := \int_0^1{\cal I}_{\rm path}(r, Q, \mu) {\rm d}r $ is the average execution price;
\item  ${\cal I}^{\infty}(Q, \mu) $ is the average price long after the end of the metaorder and can be decomposed into a predictable part ${\cal I}^\infty_{\rm info}(Q, \mu)$ and the response to the metaorder in question ${\cal I}^\infty_{\rm mec}(Q, \mu)$.
The latter is the most relevant quantity regarding price formation, since the former represents \textit{alpha} biases that has \textit{a priori} no link with the market's mechanical reaction to the order.
\ei

Note that all these quantities are implicitly averaged over all residual noise or variables.
The effect of such metaorders on the price is qualitatively well-known. While the metaorder is executed, the pressure exerted on the price tends to make it rise so ${\cal I}_{\rm path}(r, Q, \mu)$ is an increasing function of $r$.
When the metaorder is completed the {price} reverts to the \textit{permanent} level ${\cal I}^{\infty}(Q, \mu) \leq {\cal I}(Q, \mu)$. Previous empirical studies have found the \textit{peak} impact to be approximately square root of the volume and the amplitude of the \textit{post-impact decay} to be about $1/3$ on average (so that ${\cal I}^{\infty}(Q, \mu) \approx {\cal I}_{\rm exec}(Q, \mu) \approx \frac{2}{3}{\cal I}(Q,\mu)$ for square root impact, as required by equilibrium models {\rr since this ensures that the price paid for the execution is fair \emph{ex-post}}). More recent studies \citep{brokmann2014slow, gomes2015market} have shown that when subtracting the \emph{predictable part} ${\cal I}^\infty_{\rm info}$ from the price, the reversion goes all the way back to zero -- so after waiting a long enough time the \emph{mechanical impact}  ${\cal I}^\infty_{\rm mec}(Q, \mu)$ of trades on the price is zero.

One can also define the \textit{execution rate} $\mu_V$, defined as the ratio between the volume of the metaorder $Q$ and the total volume traded by the market during the same period, $V_M$,
\be
|Q| = \mu_V V_M \;.
\ee
We will thus study the peak impact ${\cal I}(Q, \mu)$, the average executed price ${\cal I}_{\rm exec}(Q, \mu)$ and the permanent impact ${\cal I}^{\infty}(Q, \mu)$, in order to identify any dependence on $T$ (or equivalently $\mu$) that goes beyond the usual rule 
of thumb~(\ref{eq:impact_emp}) which predicts impact to be independent of the execution speed. Throughout this study, price is taken to be the \emph{traded price}. Since we only consider aggressive metaorders (cf. Sec. ~\ref{sec:metaorders}), this amounts to studying the ask for buy metaorders and the bid for sell metaorders.

\subsection{Metaorder decomposition and properties}\label{sec:metaorders}

\begin{figure*}[ht!]
\centering
\includegraphics[scale=0.6]{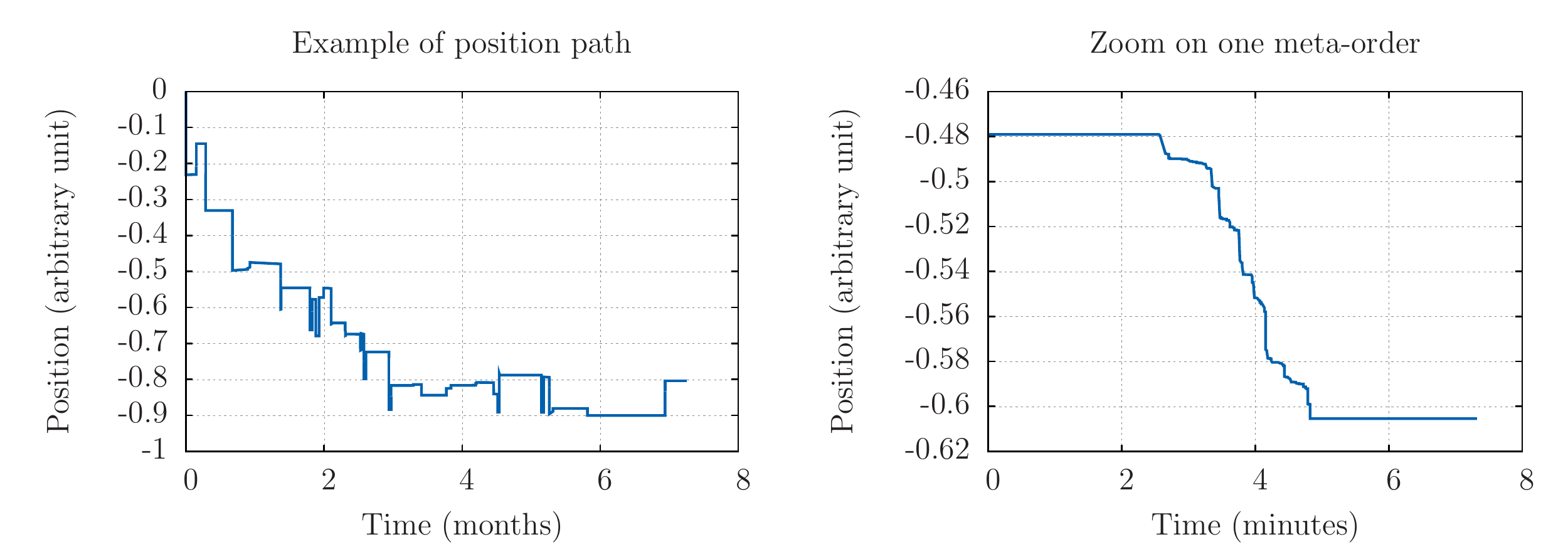}
\caption{\small{ Typical position vs time for one of the $1\%$ most active traders in terms of volume. \textit{(left)} position path during approximatively 12 months. One can see that metaorders are clearly identifiable and alternate with long periods of inactivity.
\textit{(right)} Zoom on a 2 minute sell metaorder composed by a dozen trades (zoom $1:50000$).
}
}
\label{fig_metas}
\end{figure*}

\begin{figure*}[ht!]
\centering
\includegraphics[scale=0.65]{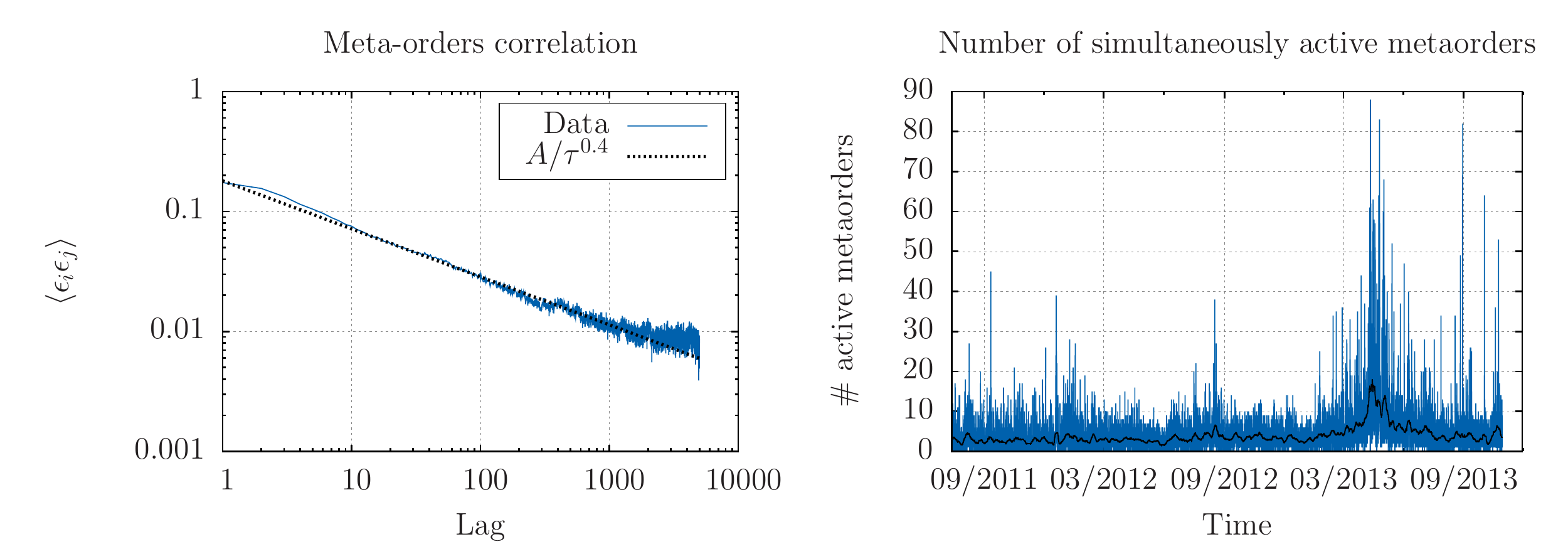}
\caption{\small{\textit{(left)} Autocorrelation function of the sign of \emph{metaorders}, {\rr ordered by starting time}. Like financial markets, not only trades but metaorders are strongly auto-correlated. \textit{(right)} Number of simultaneously active metaorders vs time. The typical value is around 5, and we can clearly observe clustering.
}
}
\label{fig_metas_vs_time}
\end{figure*}

\begin{figure*}[ht!]
\centering
\includegraphics[scale=0.65]{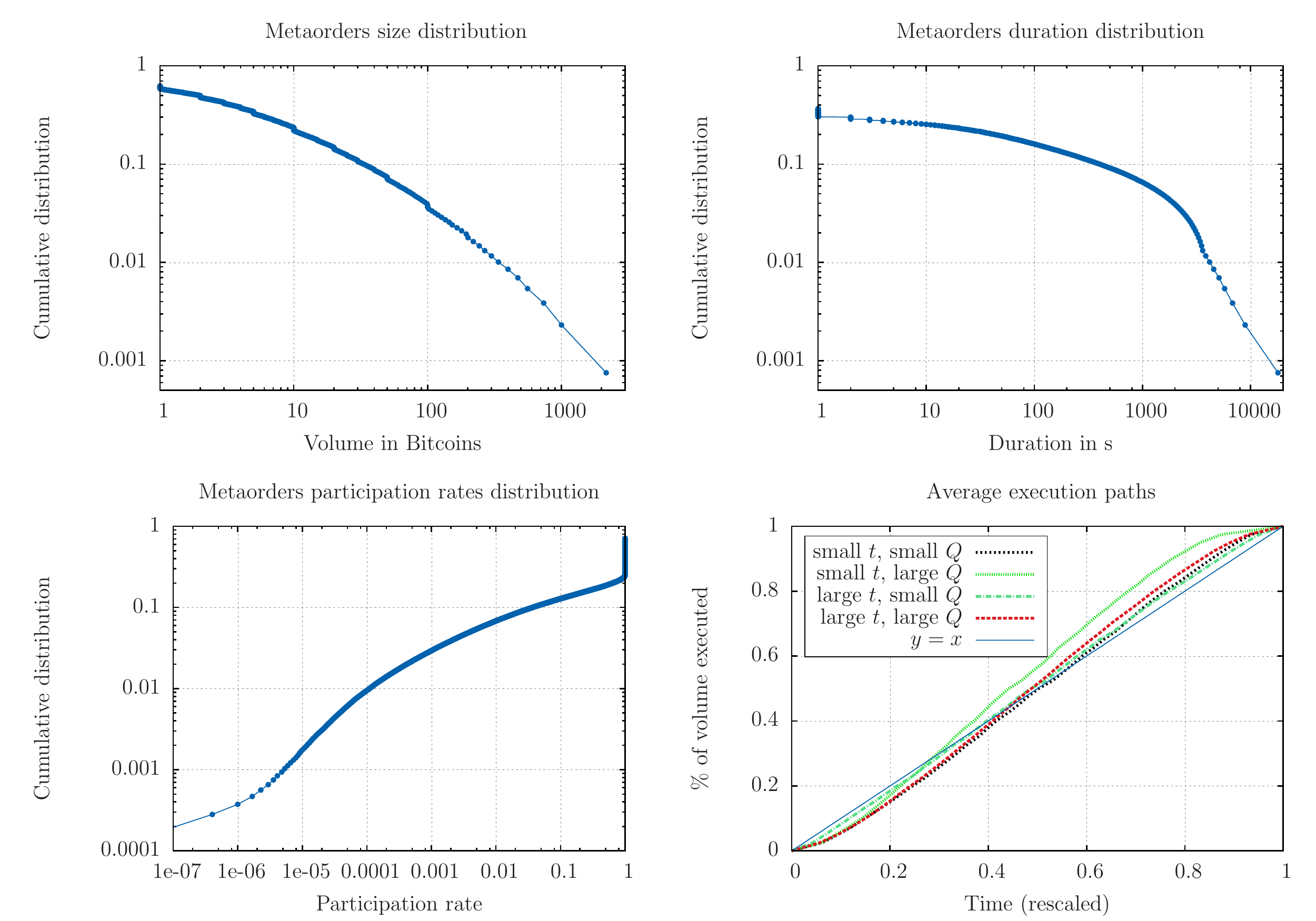}
\caption{\small{ \textit{(top and bottom left)} Metaorder size, duration and participation rate distributions for the whole market, with no clear power-law fit for any. Note that the durations are much shorter than usual metaorder durations on financial markets. \textit{(bottom right)} {\rr Percentage of volume executed vs time elapsed since the start of the metaorder}, which appears to {\rr grow roughly linearly} for all ranges of volume and duration {\rr (here the volume threshold between small and large has been fixed at $200$BTC and the time threshold at $100$s)}.
}
}
\label{fig_metas_distrib}
\end{figure*}

The first operation needed in order to study market impact is to spot the metaorders: due to the extreme irregularity and heterogeneity between the traders' typical position paths, 
usual time series decomposition methods~\citep{toth2010segmentation} are not relevant here. In order to identify these large buy and sell metaorders for this particular data, in such a way that no conditioning in the 
start/end sequences 
of the metaorders is introduced (most intuitive techniques may create mean-reversion biases), we used the following: For each trader, we defined the start of a metaorder to coincide with the 
first aggressive\footnote{We only consider aggressive orders since they are a much better proxy for system perturbations. Limit orders in the contrary may have been posted long in advance without a specific intention nor view on the price: By ignoring them we therefore limit any adverse selection that might bias our study. Besides, the fact that {\rr the execution schedule is roughly linear} strongly suggests that we do not bias the results by making this choice.} order placed after a given period of inactivity\footnote{The relevant scale for this inactivity period has been empirically determined to be about one hour.}. We define the end point of the metaorder either as the last order before a (new) inactivity period, or as the point where the 
trader trades in the opposite direction. While the introduction of a time scale to define inactivity periods could seem to be arbitrary, it is the case in practice that the metaorders are so clearly distinguishable that the sensitivity to 
this time scale is minimal (Fig.~\ref{fig_metas}): for one particular trader, metaorders are very clearly separated  from each other.
This way, from over 14 million trades we recover over 1 million metaorders of variable sizes/durations. {\rr A few percent of the trades are not assigned to a metaorder, corresponding to mean-reverting trades which by definition have a conditioning bias. Table \ref{table:n} presents the repartition of metaorders in terms of number of child orders. As expected, more than half are one-shot trades whereas $10\%$ percent are composed of more than 5 trades. However, as a result of the tick being small, more than $20\%$ of the one-shot metaorders cross several price levels.}

\begin{table}
\centering  
\begin{tabular}{|c|c|c|c|c|}\hline
$\#$ of child trades & $1$ & $2\leq n\leq4$ & $5\leq n \leq 9$ & $10 \leq n$ \\  \hline
$\%$ of metaorders & $61\%$ & $29 \%$ & $6.5\%$ & $3.5 \%$ \\  \hline
\end{tabular}
\label{table:n}
\caption{{\rr Number of child orders per metaorder.}}
\end{table}
Fig.~\ref{fig_metas_vs_time} shows the autocorrelation function of metaorder signs which is slowly decaying with a power-law exponent around $0.4$ {\rr -- which happens to be similar to the autocorrelation exponent of trades themselves --}  as well as the number of metaorders that are simultaneously active in the market at any point in time, which present clear clustering with a typical value around 5. 
The salient statistical characteristics of metaorders are presented in Fig.~\ref{fig_metas_distrib}. It presents the metaorder sizes and durations distributions, which are crucial ingredients in many market impact and price models \citep{farmer2013efficiency, donier2012market, gabaix2003theory}. Contrary to what is usually required in such models,
none of these distributions are clear power-laws -- and particularly not on \emph{all} time scales\footnote{{\rr If however one estimates a tail exponent using the Hill estimator \citep{hill1975simple} for volumes greater than $10$BTC, one finds a tail exponent of $-2$ for the probability density function.}}. This challenges such impact models, since in these models the impact function is closely related to metaorders distributions {--} and a power-law with exponent $3/2$
is required to reproduce the square-root impact{. This} is a strong hint that the distribution of order sizes is not a {fundamental input to explain the shape of market impact}, {since it is neither clearly a power-law nor universal}.
 Most importantly, {above} figure shows that the average execution profile is {\rr linear in time}. 
 This property ensures that $\mu$ is well defined in the sense that it is on average constant during execution {\rr (one also needs to check that the executed volume in the market is constant, which is the case as shown e.g. by Fig.~\ref{fig_metas_during_metas})}. This will allow us to properly compare points within impact trajectories, ${\cal I}_{\rm path}(r, Q, \mu)$, with peak impacts ${\cal I}(rQ, \mu)$.

\subsection{Conditioning checks}

\begin{figure*}[ht!]
\centering
\includegraphics[scale=0.65]{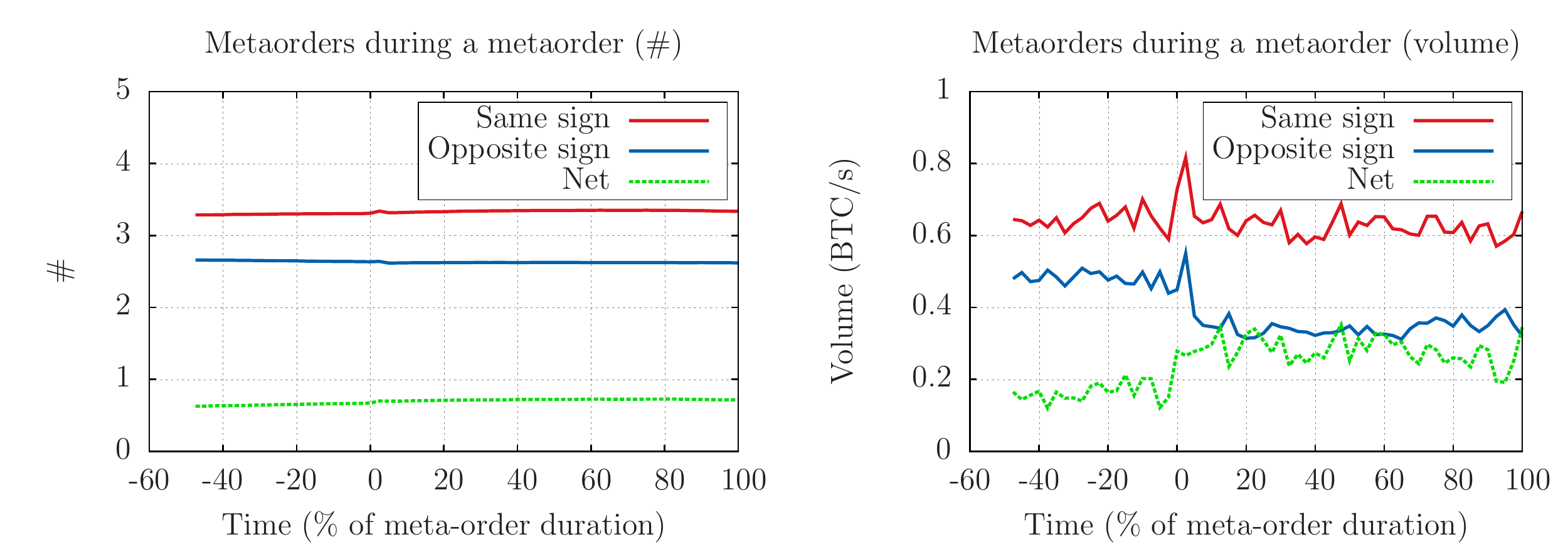}
\caption{\small{ Here we selected all metaorders with duration of approx. 2 minutes. \textit{(left)} We plotted the number of active metaorders in the same direction than the metaorder considered after subtracting it
 (green), in the opposite direction (red) and the difference (blue). \textit{(right)} The same plot, in terms of volume. On both plots, time is normalized so that the metaorders start at 0 and end at 100.
}
}
\label{fig_metas_during_metas}
\end{figure*}

 It has been shown in the previous section that average execution style is {\rr linear in time}, 
so that we can really study the effect on the price of metaorders with constant execution speed on average\footnote{as opposed e.g. to~\citet{Farmer:new} in which they find strongly front-loaded executions.}. 
In this section, we proceed to some other checks concerning the behaviour of the rest of the market while metaorders are being executed. Indeed, {the usual argument against the square-root law assumes them to be conditioned to the rest 
of the market in such a way, that the apparent concavity of the marginal impact is only an artefact.} In particular, anticipating Sec.~\ref{sec:further}, we look into the buy/sell market order flow while a metaorder is executed, to check whether dynamical effects exist
in order flow that could result in concavity: for example, a higher correlation with market imbalance at the beginning of the metaorder would result in sharper price changes at the start of the trajectories\footnote{
This would imply some sort of synchronization between the agents, either exogenous 
 (e.g. the agents react to the same news thus starting their metaorders at the same time, but some stopping before the others) or endogenous (e.g. arbitrageurs are able to detect metaorders and push the price up at its beginning to sell
 back later when the price has risen).}. {\rr Since we have the start and end times of every metaorders as well as their execution rates, we can compute at any point in time the number/volume of active buy/sell metaorders in the market. In Fig.~\ref{fig_metas_during_metas} we compute these quantities during the execution of chosen metaorders (here all metaorders of duration about 2 minutes) and plot the average paths over all these metaorders to look for such dynamical effects.}
The data shows that such a synchronisation of metaorders is almost non-existent, either in terms of number of metaorders or in terms of volume, so that concavity of impact is not generated by some kind of synchronization between the agents\footnote{One can note the
interesting (unrelated) fact that market order volume in the opposite direction slightly decreases while a metaorder is being executed. {\rr Also, during a buy (resp. sell) metaorder the other metaorders present on the market tend to go in the same direction. This confirms the fact that metaorders are ``informed'' on average. A more in-depth discussion on this notion of information can be found in \citet{donier2015walras}.}}. Since 
the study of impact paths in Sec.~\ref{sec:further} will show that the {square-root} law holds for {the whole} \emph{trajectories} {and not only for the peak impact}, one can assert that concavity is \emph{true} in the sense that it does not stem from conditioning but is really the way the market absorbs an excess of supply or demand.

\section{The square root impact law for the Bitcoin/USD exchange market}\label{sec:impact}

\subsection{The square root law of peak impact for individual metaorders}

\begin{figure*}[ht!]
\centering
\includegraphics[scale=0.66]{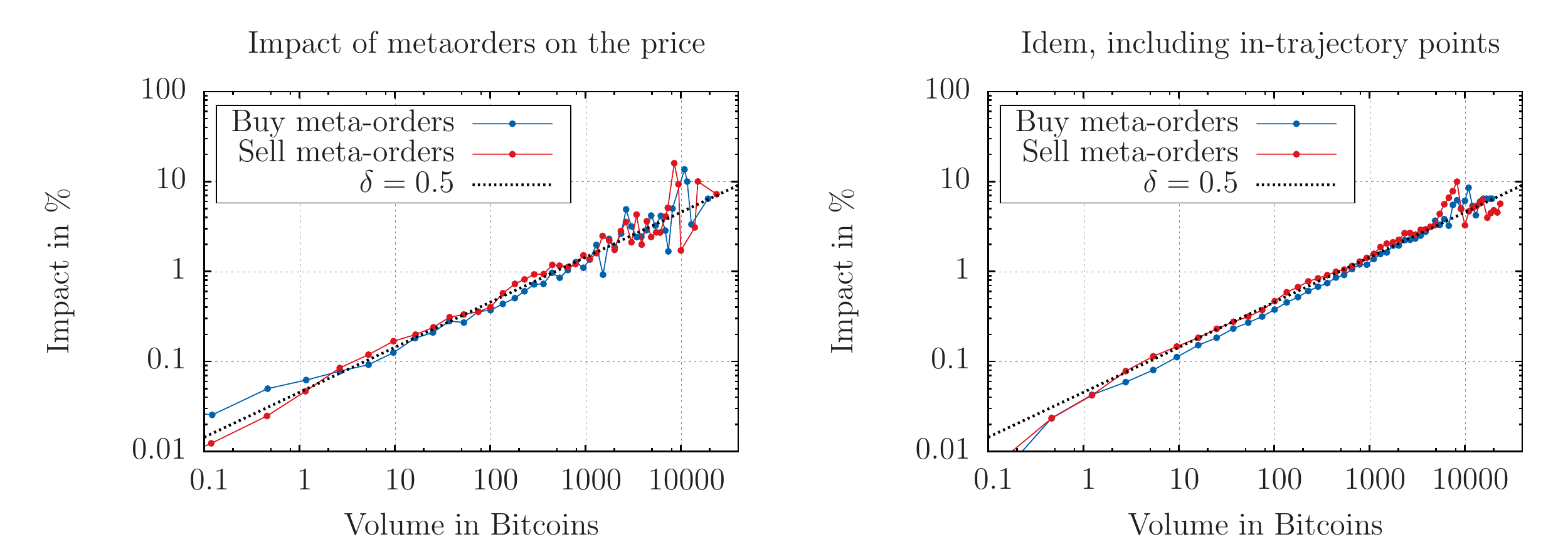}
\caption{\small{Market impact $\langle {\cal I}(Q, \mu) \rangle_{\mu}$ (averaged over all execution rates $\mu$), follows the same square root law {as} is observed by banks and hedge funds on financial markets (plots in log-log scale). {\rr Each point represents the average impact of all metaorders in a given range of volume.} The impact exponent $\delta$ is found to be very close to $0.5$, and the $Y$-ratio is around $0.9$. One should emphasize that this power-law behaviour appears at the smallest scales and holds over 4 decades. 
\textit{(left)} Only end points of metaorders ($\sim$1M data). \textit{(right)} 41 point per metaorder (every 2.5\% quantile of volume), giving 27M data points. Part of these points being degenerate, one can assess the number of effective points around 
a few millions.
}
}
\label{fig_impact}
\end{figure*}

For each of the 1M metaorders we identified above, we considered the impact defined as 
\be
\mathcal I (Q, \mu)= {\cal I}_{\rm path}(r=1, Q, \mu),
\ee
i.e. the difference between the first and the last executed price, quantifying the reaction of the market to the trader's order. Note that here impact is measured as the peak price (with the initial price gauged to zero).
 The result is shown in Fig.~\ref{fig_impact}: In spite of the very special features of the Bitcoin market, a concave impact law (depicted as a straight line)  
\begin{equation}
\langle {\cal I}(Q, \mu) \rangle_{\mu} \approx \widetilde{Y}Q^{\delta}\;,
\end{equation}
fits the data points very well from the smallest scales and over 4 decades with $\delta \approx 0.5$ and $\widetilde{Y} \approx 4.5\cdot 10^{-2}$. Normalizing by Bitcoin average volatility and daily volume gives a \emph{Y-ratio} (as defined in Eq.~\ref{eq:impact_emp}) of $Y \approx 0.9$,  close to the value reported on ``mature'' financial markets, e.g. futures or stocks \citep{toth2011anomalous,brokmann2014slow}. {For a more in-depth study of the \emph{Y-ratio}, see Sec. ~\ref{sec:yratio} below.}
Thus, peak impact for the Bitcoin is consistent with the square-root law despite the prototypical nature of the Bitcoin market. This confirms the view expressed in~\citet{toth2011anomalous} and \citet{mastromatteo2014agent} that market impact depends neither on microstructure (with e.g. arbitrage and high-frequency trading), nor on a clear metaorder distribution. As such, it directly challenges the explanation of concave market impact in terms of rational equilibrium theories.

\subsection{Square root impact trajectories}
 
\begin{figure*}[ht!]
\centering
\includegraphics[scale=0.66]{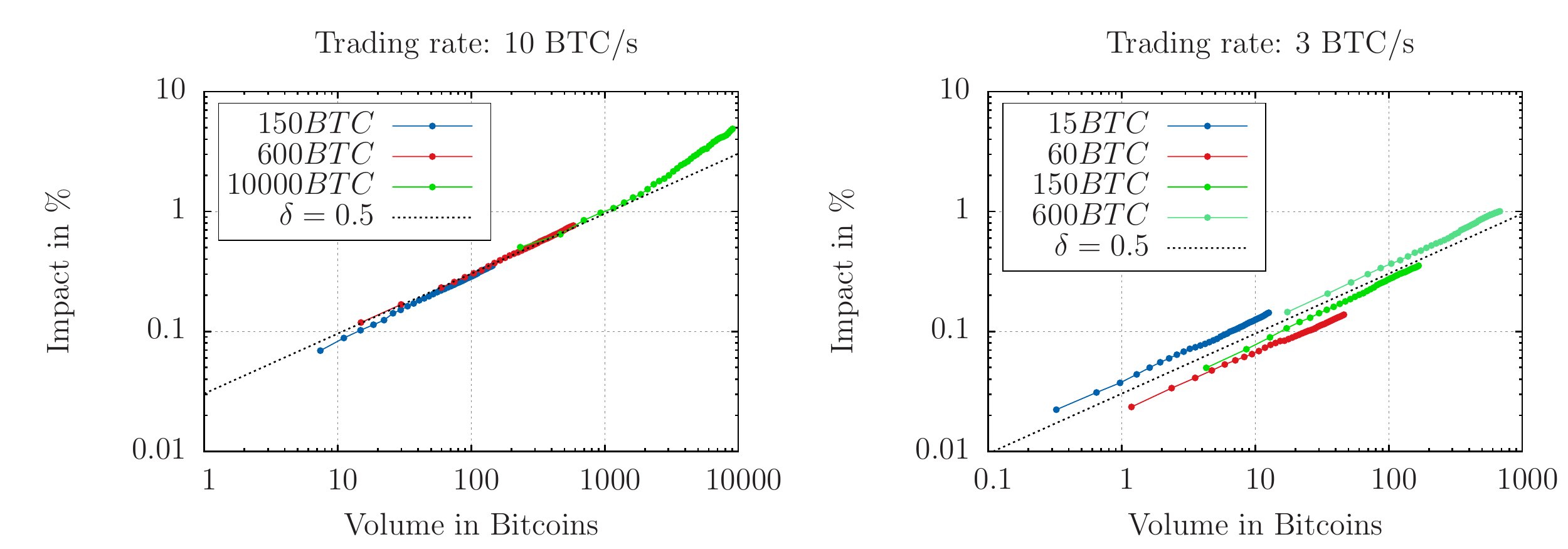}
\caption{\small{ Impact paths ${\cal I}^{\rm path}(p, Q, \mu)$ in decimal loglog plot, {\rr for different metaorder volumes (cf. legends), for} \textit{(left)} $\mu = 10 BTC/s$  
and \textit{(right)} $\mu = 3 BTC/s$, and for $r\in [0,1]$ for each couple ($Q$, $\mu$). The first value has intentionally been chosen high, so that it survives the criticism raised in Section~\ref{sec:speed} that on average other metaorders in the same direction  are observed, which tends to artificially increase impact measures. One can observe a liquidity breakdown leading to an asymptotically linear impact when important pressures are maintained for too long on the same side of the order book.
}
}
\label{fig_path}
\end{figure*}

We now turn to the study of impact trajectories, i.e. the quantity ${\cal I}_{\rm path}(r, Q, \mu)$ for given $Q$, $\mu$ and $r$ varying from 0 to 1. Unless mentioned otherwise, we average over all other quantities in the following (like e.g. daily volatility, daily traded volume etc.). Fig.~\ref{fig_path} shows the results, putting into light 
two facts of particular interest. The first is the answer to question whether the impact trajectories follows the same law as peak impacts.
On the Bitcoin, where execution paths are {\rr roughly executed linearly in time}, the agreement is remarkable, meaning that while the metaorder is not finished the market makes no difference between a metaorder that will stop 
soon and a metaorder that will continue\footnote{This is at variance with \citet{Farmer:new} where average executions are not {\rr linear in time} but rather front-loaded execution, resulting in very particular price trajectories.}. We find empirically that
\be
{\cal I}_{\rm path}(r, Q, \mu) = {\cal I}(rQ, \mu) \;.
\ee
Hence, if $t$ is the time elapsed since the start of the execution, {\rr and since $\mu$ is relatively constant for the metaorders in our dataset,} one can write impact as 
\be
{\cal I}_{\rm path}(r, Q, \mu) ={\cal I}(t, \mu) = f(\mu)t^{\delta} (:= \widetilde{f}(\mu)(\mu t)^{\delta})\\
\ee
where $\delta \approx 1/2$. {\rr This in particular allows one to include in-trajectory points in the measurement of peak impact, cf. Fig.~\ref{fig_impact} (right) as they would have been the peak impact of the metaorder, had it stopped before. Note that in the above equation we implicitly average over any other variable so that $f$ only depends on $\mu$ (with an assumption of independence).}

{\rr Second, one observes that the value of the $\widetilde{Y}$-factor appears to be noisy as impact lines do not superimpose, but are all perfectly parallel, across different $Q$’s.
A possible cause is the time variations of $\widetilde{Y}$ during the period considered (see Fig.~\ref{fig_path}), leading to a conditioning effect as soon as the distributions of  $Q$ is not similar during high- and low-$\widetilde Y$ periods. This suggests that the price during a metaorder execution can be written as
\be\label{eq:var_factorization}
I(t, \mu) = \widetilde{Y}(t)\sqrt{\mu t} + \sigma W_t\;,
\ee
where $\widetilde{Y}(t)$ accounts for a (slowly) time-varying liquidity~\footnote{By writing this integrated form one implicitly assumes that $\widetilde Y(t)$ is constant throughout the execution and varies at scales that are slower.} and $\sigma$ represent some additional market noise.}

\subsection{The ``{Y-ratio}''}\label{sec:yratio}

\begin{figure*}[ht!]
\centering
\includegraphics[scale=0.55]{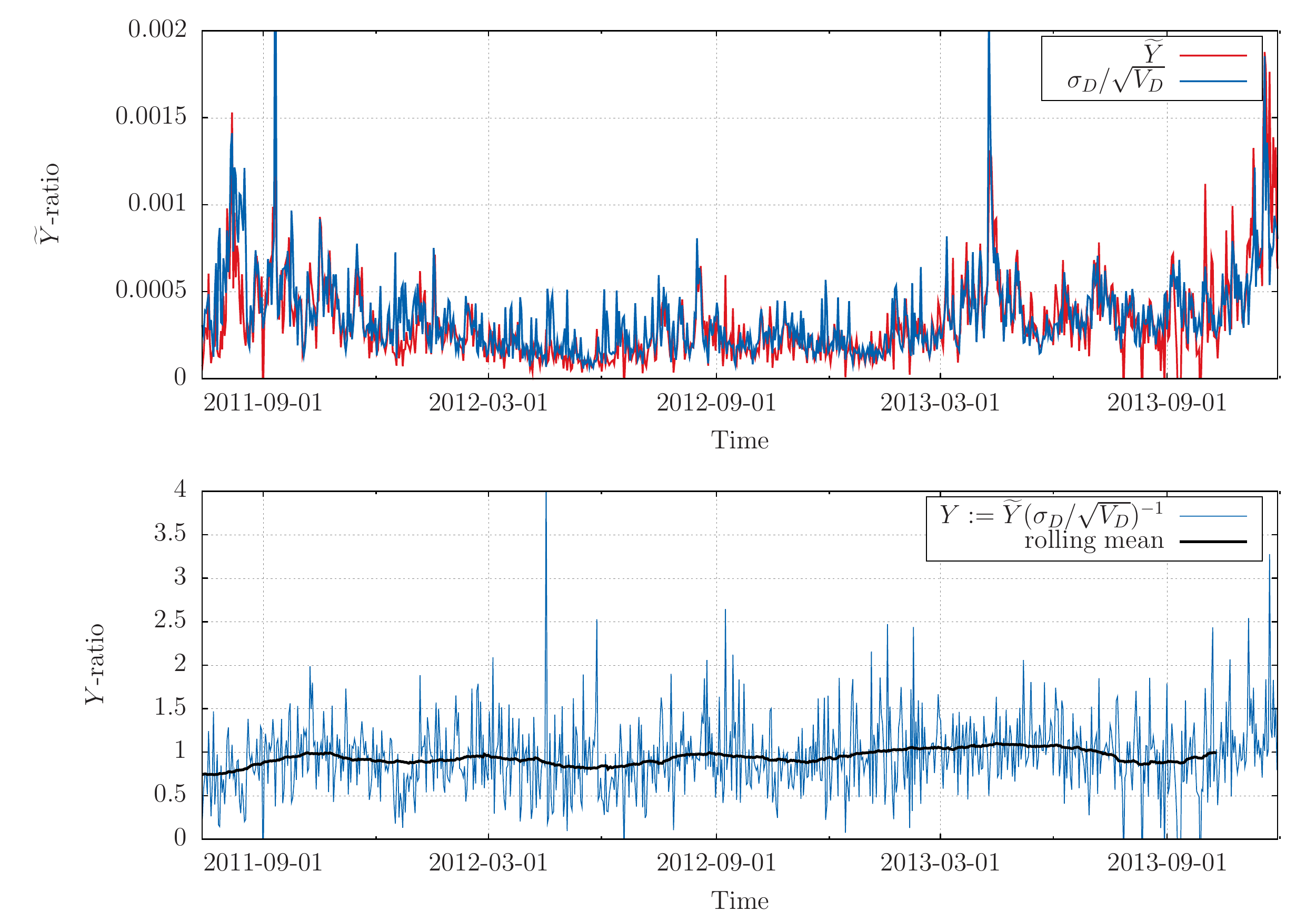}
\caption{\small{\textit{(top)} Raw impact pre-factor $\widetilde{Y}$ vs time. We also plot the usual normalization $\sigma_D\sqrt{V_D}$ to show that it accounts for the major part of the non-stationariness, particularly during extreme market events (e.g. April 10, 2013 major crash). \textit{(bottom)} \emph{Y-ratio} as defined in Eq. ~\ref{eq:impact_emp}, which oscillates around its mean value $Y_0 \simeq 0.9$.
}
}
\label{fig:yratio}
\end{figure*}

\begin{figure*}[ht!]
\centering
\includegraphics[scale=0.75]{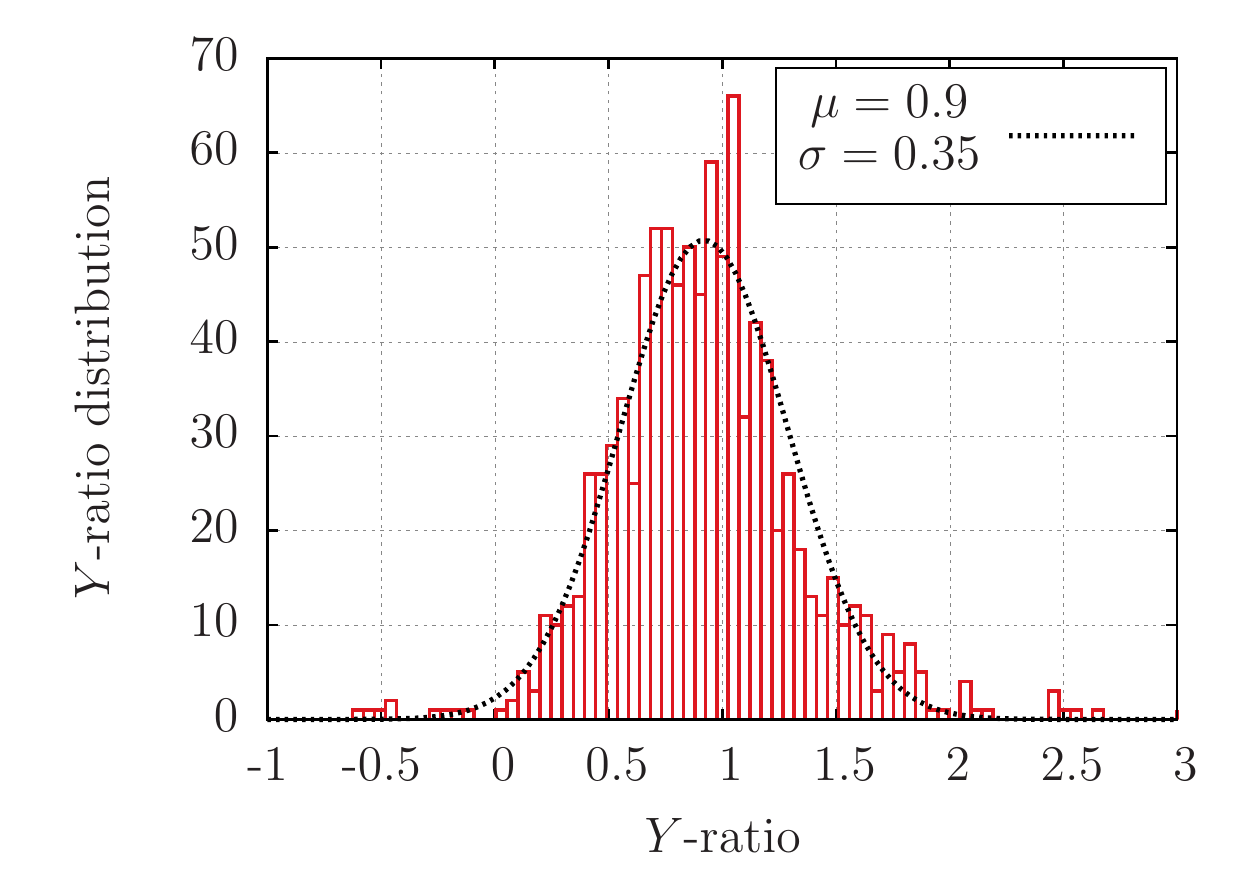}
\caption{\small{\emph{Y-ratio} distribution and its Gaussian approximation ${\cal N }(Y_0, \Sigma_Y)$ with mean $Y_0 =0.9$ and $\Sigma_Y = 0.35$.
}
}
\label{fig:yratio_distrib}
\end{figure*}

Until now, much emphasis has been put in the literature on the study of the dependence in $Q$ but very few studies have been realized on the pre-factor $\widetilde{Y}$ -- or equivalently the \emph{Y-ratio} defined in Eq. ~\ref{eq:impact_emp}. We devote this section to a temporal analysis of both these pre-factors.
For each metaorder $i$, we compute its individual $\widetilde{Y}_i$ as the ratio ${\cal I}(Q_i)/\sqrt{\mid Q_i \mid} \times \text{sign}(Q_i)$. For each day, we compute the daily average $\widetilde{Y}$ as a volume-weighted average of all individual ratios. In parallel, we compute the daily realized volatility $\sigma_D$ and traded volume $V_D$, and we compare the daily $\widetilde{Y}$'s to the corresponding $\sigma_D / \sqrt{V_D}$ ratios for every day in our dataset. For each day, we also plot the actual \emph{Y-ratio}. Results are presented on Fig. ~\ref{fig:yratio} and show that after such rescaling, the \emph{Y-ratio} becomes nearly time independent. Its distribution is plotted on Fig. ~\ref{fig:yratio_distrib} and is well approximated by a Gaussian distribution ${\cal N }(Y_0, \Sigma_Y)$ with mean $Y_0 =0.9$ and standard deviation $\Sigma_Y = 0.35$.


From these results, we can draw two conclusions of particular interest. First, it validates the scaling form of Eq. ~\ref{eq:impact_emp} proposed in \citet{Barra:1997, grinold2000active, toth2011anomalous}. Indeed, the non-stationariness of the impact pre-factor $\widetilde{Y}$ is well encoded in the ratio $\sigma_D\sqrt{V_D}$. Besides, the residual a-dimensional \emph{Y-ratio} is shown to be of order unity with a standard deviation of the same order of magnitude so that it essentially lies in the interval $[0,2]$. In this light, Eqs. ~\ref{eq:impact_emp} and  ~\ref{eq:var_factorization} can be merged together so that impact reads
\be\label{eq:var_factorization_sto}
I(t, \mu) = \sigma \left[(Y_0 + \sigma_Y \eta )\sqrt{\frac{\mu t}{V_D}} +  W_t \right]\;,
\ee
{\rr where $\sigma_Y$ accounts for some possible noise on the value of $Y_0$.} The relationship between the \emph{liquidity noise} $\sigma_Y$ and the \emph{market noise} $\sigma$\footnote{Both of which contribute to the $\Sigma_Y$ evoked above.} -- and their dynamics -- have not been investigated here, although it is a topic of interest for further studies.

The other conclusion is more macroscopic and relates to the study of market stability. Indeed, the scaling form for $\widetilde{Y}$ holds particularly well during extreme market events, such as the major crash that occurred on April 10, 2013, {\rr as evidenced in \citet{donier2015markets} who show that the impact pre-factor $\widetilde{Y}$ is a relevant proxy for market liquidity, even at scales much larger than that of impact.} This definitely relates the microscopic aspect of price formation to its macroscopic characteristics such as its propensity to crash -- see \citet{kyle2} for similar discussions on financial markets. In this light, the understanding of how trades impact prices appears more crucial than ever to understand, detect -- and perhaps even control? -- market instabilities.

\section{Impact, execution speed and correlations with the order flow}\label{sec:further}

\subsection{Impact trajectories of the bid and the ask}

\begin{figure*}[ht!]
\centering
\includegraphics[scale=0.55]{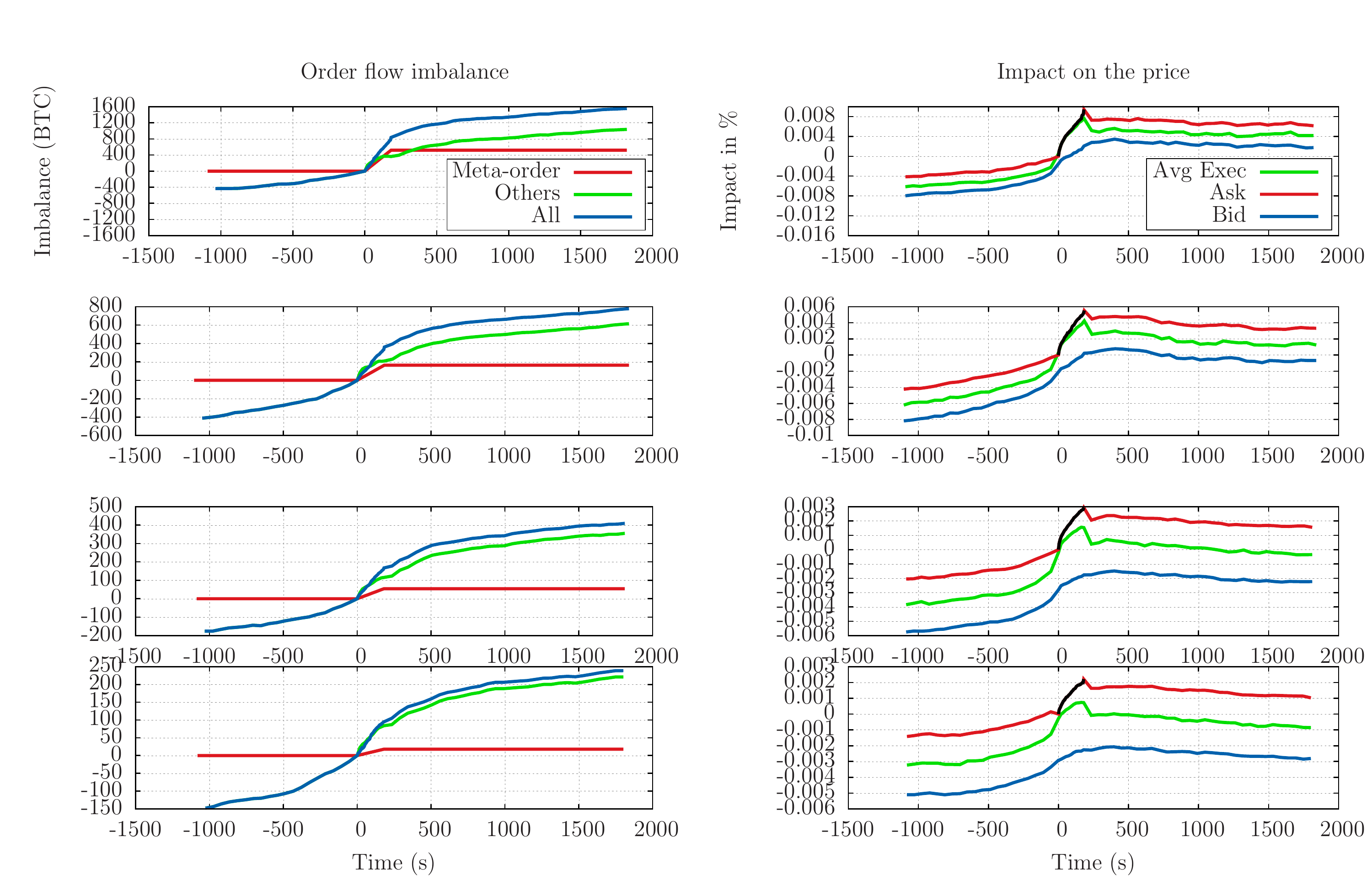}
\caption{\small{{\rr (left) Order flow imbalances measured as the cumulated volumes of signed market orders during the execution of metaorders, for different metaorder volumes. For each panel, the metaorder starts at time 0 as shown by its executed volume in red. The blue line is the signed cumulated order flow of the whole market, whereas the green curve subtracts the metaorder from it to keep only the residual order flow.} (right) Corresponding impact paths. In black, the actual execution path. In red, the ask path, and in light blue, the bid path. In green, we also plot the average execution price of the market. One can see that even for large volumes, impact is at most 1 or 2 spreads, not mentioning the fees that are of same order of magnitude than the spread: impact costs are dominated by friction costs, challenging martingale and fair pricing conditions.
}
}
\label{fig_impact_examples}
\end{figure*}
\begin{figure*}[ht!]
\centering
\includegraphics[scale=0.55]{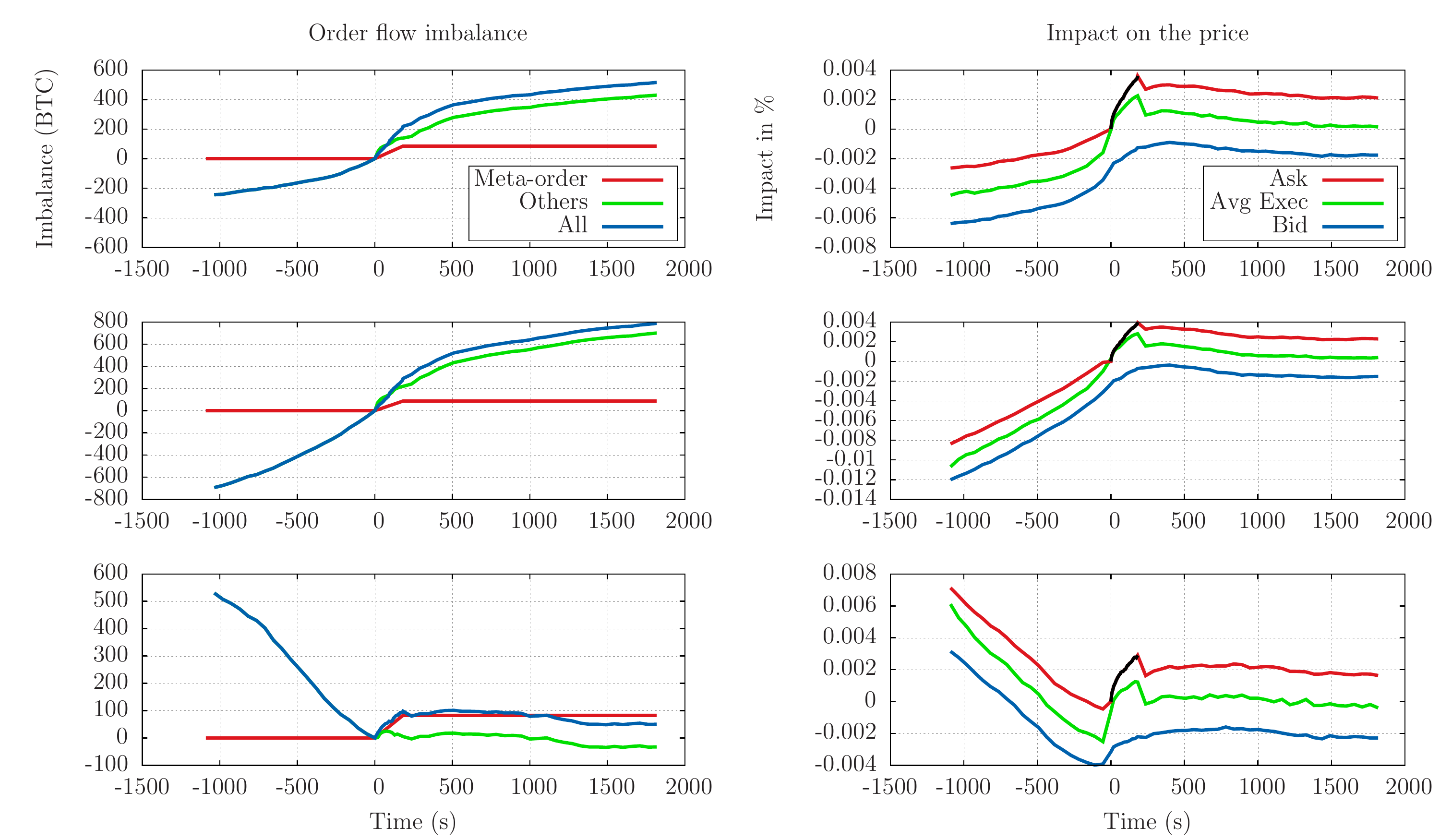}
\caption{\small{Same pictures as in Fig.~\ref{fig_impact_examples} for various types of conditioning \textit{(top)} Unconditioned metaorders. \textit{(middle)} Trending
metaorders. \textit{(bottom)} Mean-reverting metaorders.
}
}
\label{fig_impact_examples_2}
\end{figure*}

The large fees on Bitcoin allow for a separate study of the bid and the ask during the execution and prevents the data to be too noisy due to high-frequency arbitrage. For the sake of simplicity, we assume that metaorders are buy orders, so that the ask 
denotes the opposite side of the order book, on which the trader executes his/her metaorder. The facts of interest are the following :
\begin{itemize}
\item Before the execution, the spread is roughly constant, and the execution direction is on average positively correlated with both the bid and the ask evolution before the execution (Fig.~\ref{fig_impact_examples}).
\item During the execution, the ask rises sharply -- following the same square root {as} the execution price -- whereas the bid follows more linearly. This is very reminiscent of what is found theoretically in~\citet{DonierBonart14}.
\item After this quick reversion, the bid and the ask remain roughly constant at a {non-zero} permanent level, in spite of the order flow pressure from the rest of the market that continues some time after the metaorder.
\item These observations hold when we condition the metaorder to be trend-following or mean-reverting, as shown on Fig.~\ref{fig_impact_examples_2}. In particular, the fact that impact is still square root at small scales for trend-following metaorders is non-trivial and reveals that \emph{trading speed matters}\footnote{It contradicts in particular equilibrium models that only consider the aggregate order flow as a relevant quantity, since in these models \emph{trading speed} is transparent.}.
\item In any case, impact does not exceed very much the order of magnitude of the spread and the fees, which questions any interpretation based on ``fair-price'' arguments, all the more so that impact is square root even at the smallest scales.\\
\end{itemize}
{\rr Consistent with the observations of Fig.~\ref{fig_metas_during_metas}, one observes that on average metaorders are positively correlated with the remaining order flow. We will see below that this plays a crucial role in the fact that the permanent component of impact is non-zero in these plots.}

{\rr Finally, one notices that the market VWAP in green lies around the mid-price in the absence of the metaorder, and is naturally biased towards the execution side during the execution -- all the more so that the execution is aggressive.}

\subsection{Impact and execution speed}\label{sec:speed}

The question of dependence of impact on the execution speed $\mu$ has been addressed very recently by \citet{Farmer:new} even though practitioners have been looking into it for at least a few years.
In this section we study the \emph{impact surface} described by the bivariate function ${\cal I}_{\rm exec}(Q,\mu_V)$
\footnote{We prefer here the average quantity ${\cal I}_{\rm exec}(Q,\mu_V)$ to the peak impact $\mathcal I(Q,\mu_V)$ since latter is much noisier. 
The mean execution price is anyway far more relevant in practice since it is directly related to \textit{execution costs}. We also preferred the execution rate $\mu_V$ to the execution speed $\mu$ as a fundamental variable since for this particular study this quantity is more relevant and yields cleaner pictures.} 
For each fixed $\mu_V$ the dependence on $Q$ is found to be square root:
\be
\langle\mathcal I^{\rm exec} (Q, \mu_V)|\mu_V\rangle \sim \sqrt{Q}\;.
\ee
\begin{figure*}[ht!]
\centering
\includegraphics[scale=0.37]{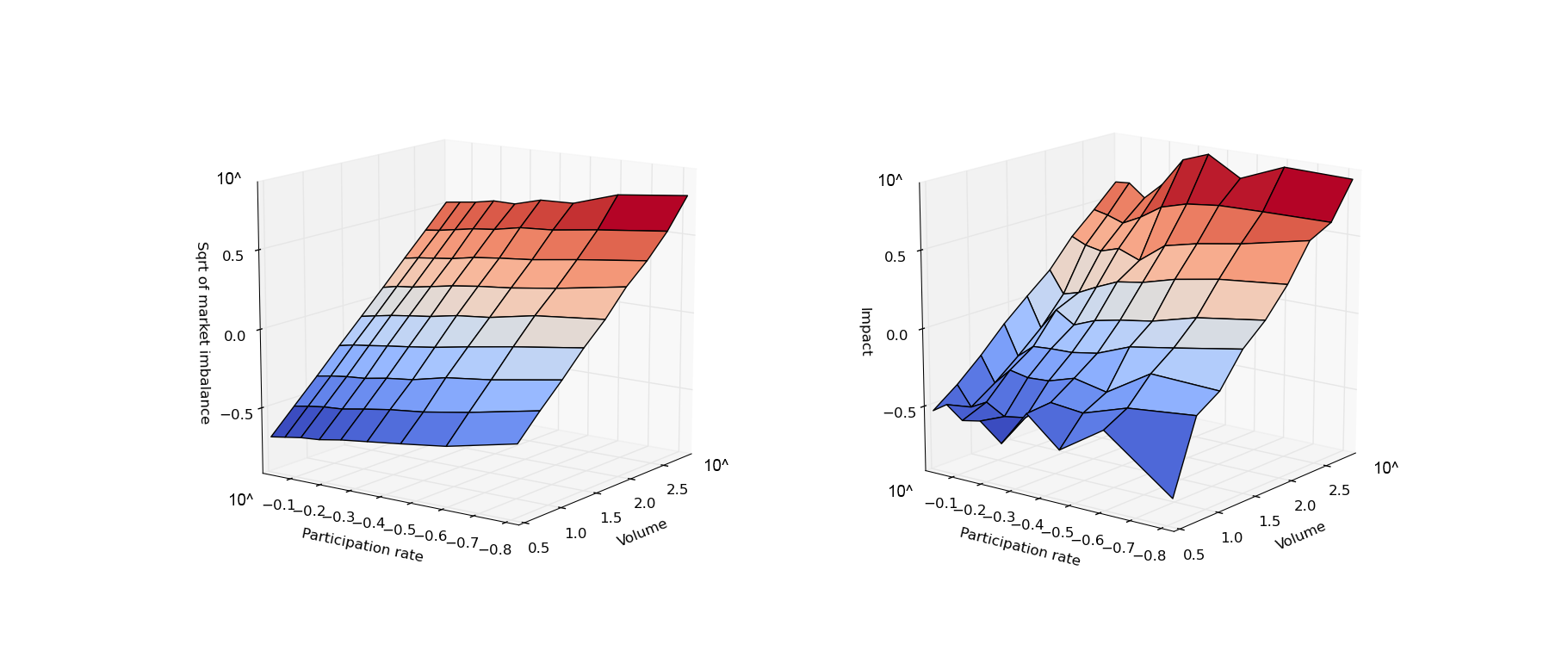}
\includegraphics[scale=0.47]{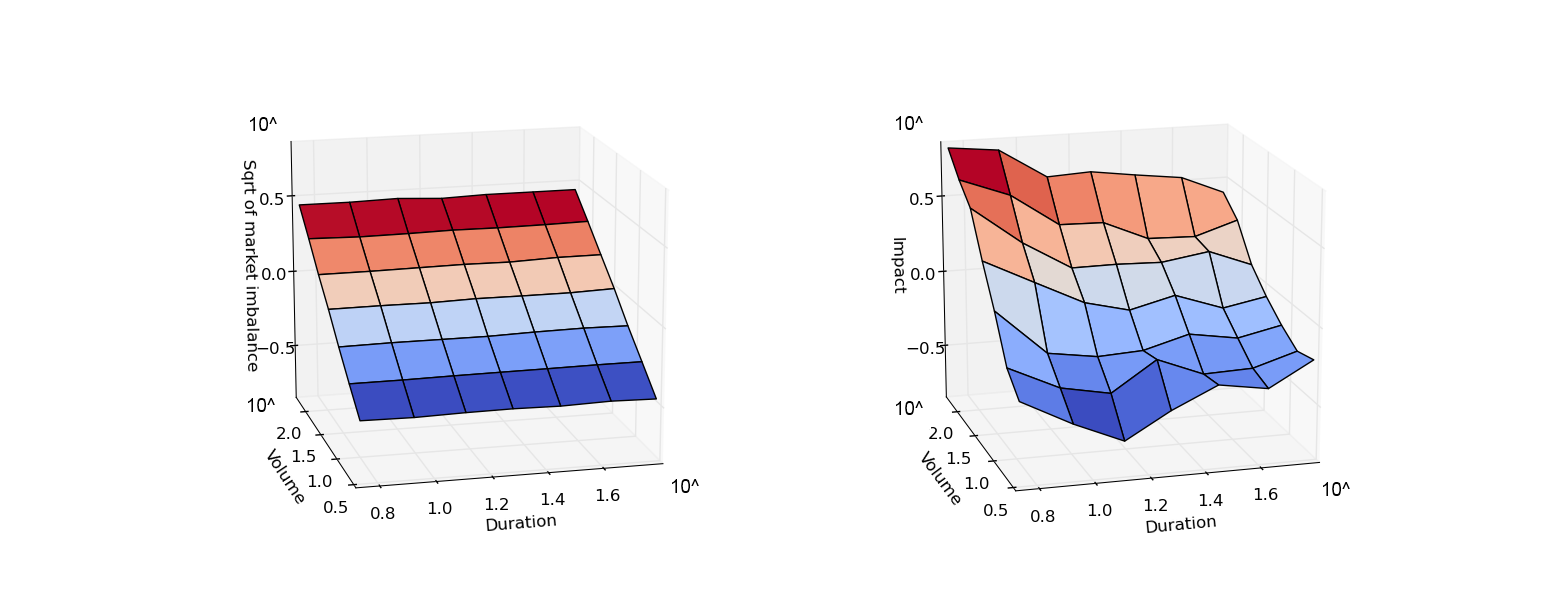}
\caption{\small{The impact surface (in log-log-log scale) for typical metaorders (top) which are correlated with the total order flow. During the execution span of the metaorder, the market reacts rather to total market imbalance 
as to the individual market order: For all participation rates market impact is almost perfectly proportional to the square root of the total order flow as defined in the main text. 
Bottom images: Impact of an isolated metaorder (for which the residual market order flow remains neutral). The total market 
imbalance corresponds to the volume of the metaorder; hence, the participation rate is not a suitable measure. Rather, the execution time should be regarded. The figure clearly displays an impact that decreases when the execution time 
increases.
}
}
\label{fig_surface}
\end{figure*}
However, the dependence on $\mu$ is somewhat surprising. First, for very high participation rates (close to $100\%$ of the volume during the period), the impact becomes unusually high. This is not surprising since the market breaks down in this regime
\footnote{this fact is of importance since such an observation would probably be impossible on financial markets, where participation rates rarely exceed $20-30\%$ -- for this very reason.}. 
More importantly however, while one could expect that impact monotonically decreases as the execution rate decreases, the observed impact actually \textit{increases} again (cf. Fig.~\ref{fig_surface}), at variance with intuition and previous 
findings on financial markets \citep{Farmer:new}. The best fit to the empirical data reads $I^{\rm exec}(Q, \mu_V) \sim Q^{\delta}/\mu_V^{\delta'}$, with $\delta\approx 0.5$ and $\delta' \approx 0.4$ (cf. Fig.~\ref{fig_surface}). The slower the execution, the larger the measured impact: this strange dependence on $\mu_V$ stresses the difference between mechanical and informational impact. Clearly, a slow execution gives other market participants the opportunity to detect the same signal and the information content of the metaorder realizes during its execution. On the other hand, fast execution leads mostly to mechanical impact as the \emph{alpha} realizes itself afterwards. 
Following this first non-intuitive discovery -- which is very specific to the Bitcoin --, for each of the data points of Fig.~\ref{fig_surface} we plotted the whole market order imbalance $ {\rm sign}(Q)\cdot V_M^{\rm signed}$\footnote{$V_M^{\rm signed} = \sum_i v_i\epsilon_i$ where $\epsilon_i = \pm 1$ according to 
whether the trade is triggered by the buyer (resp. seller) and $v_i$ is its volume.}. The similarity of both pictures (see Fig.~\ref{fig_surface} top images) is striking: during metaorders, 
impact is a very nice square root of \textit{global market imbalance}. This leads to the following conclusion: Market impact is not a reaction to \textit{individual} metaorders, but to the \textit{whole} order flow. This seems rather natural since orders are anonymous, hence the aggregated order flow 
should be the only relevant quantity. 
 
\subsection{Permanent impact and correlation}
 
\begin{figure*}[ht!]
\centering
\includegraphics[scale=0.75]{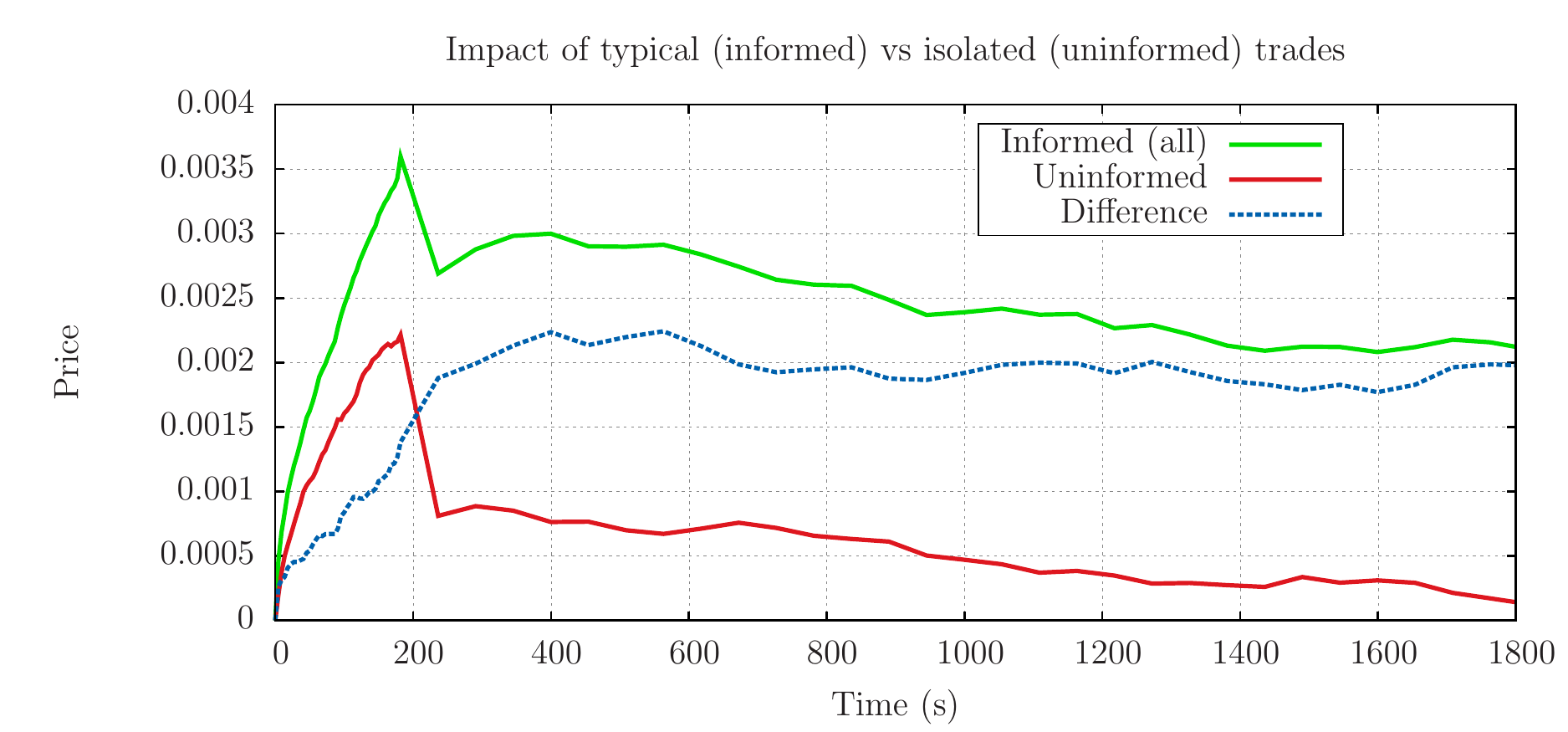}
\caption{\small{Impact of ``informed'' metaorders vs. ``uninformed'' metaorders (i.e. isolated metaorders) {\rr on the opposite best price}. While the former have a permanent impact due to their correlation with the residual order flow, the latter do not affect the price in the long run -- or very few.
}
}
\label{fig_uninformed}
\end{figure*}

One question remains however: How to study the mechanical impact of \textit{one isolated} metaorder? We can answer this question by searching the data for metaorders that are not correlated with the rest of the market, i.e. in the course of which the residual order flow does not trend nor anti-trend: we selected all metaorders that account for more than 75\% of the market net imbalance on $[t_0, t_0 + 10T]$ where $t_0$ is the starting time of the metaorder and $T$ is its duration, so that the rest of the market can be considered neutral during the measurements {\rr -- hence the terminology of \textit{isolated} metaorder}. Fig.~\ref{fig_uninformed} compares the impact of randomly chosen metaorders (which are positively correlated to the rest of the market), that we will refer to as ``informed'', and isolated metaorders (that we will refer to as ``uninformed'') and clearly shows that impact of isolated metaorders decreases far below the $2/3$ threshold. {\rr Note that the above selection labels about $3\%$ of the metaorders as isolated.}
This illustrates the fact that impact is built up by an ``informational'' component, that reveals itself during and after the execution and results in an apparent permanent impact\footnote{ This is Hasbrouck's~\citep{Hasbrouck91} definition of ``information''.} (in the case of 
isolated orders this component is by definition zero or very small) and a ``mechanical'' component whose shape after execution is consistent with a decay all the way to zero, {\rr in agreement with the findings of \citet{gomes2015market}}. This 
suggests that the mechanical peak impact is found by removing the permanent part of the impact: 
\be
{\cal I}^{\infty}_{\rm mec}(Q,\mu) \approx \mathcal I(Q,\mu) - \mathcal I^{\infty}(Q,\mu)\;.
\ee
One might ask whether the ``informed/uninformed'' terminology is appropriate here, since we never wonder about external information. Actually, the fact that isolated metaorders have no permanent impact, and therefore no ``informational content'' on average, indicates that the permanent component of the impact -- that one may refer to as the \emph{informational content} of the metaorder -- should be interpreted as a \emph{correlation} between the metaorder and future traders' behaviours and not as some information about any ``fundamental price'' (see \citet{donier2015walras}).
Concerning the dependence of impact on the execution speed, the same kind of impact surface can be plotted after conditioning on isolated metaorders only, showing that after accounting for the bias the impact actually \emph{decreases}, as expected intuitively, when the execution speed
decreases:
\begin{equation*}
  \mu_1<\mu_2 \Rightarrow {\cal I}^{\infty}_{\rm mec}(Q,\mu_1) < {\cal I}^{\infty}_{\rm mec}(Q,\mu_2)\;.
\end{equation*}
 However, the measure is too noisy to quantify precisely this dependence.
Such a behaviour is strikingly reminiscent of what is obtained within the reaction-diffusion framework -- and less consistent with equilibrium models which so far do not deal with the issue of execution speed.

\section{Summary of main results}\label{sec:sum}

We have presented a very detailed analysis of market impact on the Bitcoin exchange market. For the sake of a clear understanding let us summarize here our main 
results:

\begin{enumerate}
  \item Large orders are split into small trades which confirms the wide-spread belief that large orders have to be executed incrementally. On our MtGox dataset over 1 million metaorders are clearly identifiable, corresponding to 14M trades.
  \item Metaorders size and duration distributions are not power-law. However, on Bitcoin they have the very nice property of being executed at a constant average rate.
  \item The market impact of a metaorder executed incrementally and {\rr linearly in time} can be reduced in a dependence on the execution speed and on the time elapsed since the start of the execution and written 
  ${\cal I}_{\rm path}(r, Q, \mu) ={\cal I}(t, \mu)$.  The square-root law is clearly confirmed as it even holds \emph{trajectory-wise} and is exact even at the smallest scales, which once again invalidates a broad class of equilibrium theories. Besides it has been shown that this cannot be explained by biases: the square-root law describes how market digests local excesses in supply /demand.
  \item { The impact pre-factor $\widetilde{Y}$ is subject to large fluctuations, that are well explained by the ratio $\sigma_D/\sqrt{V_D}$ where $\sigma_D$ is the daily volatility and $V_D$ the daily volume. The residual \emph{Y-ratio}, as defined in Eq. ~\ref{eq:impact_emp}, fluctuates around a mean-value of order unity: to be more precise, its distribution is well approximated by a Gaussian distribution of mean $Y_0 = 0.9$ and standard deviation $\Sigma_Y = 0.35$. The full stochastic impact formula we propose is thus given by Eq. ~\ref{eq:var_factorization_sto}.}
  \item We have presented strong evidence that the market reacts to the total order flow and not to distinct metaorders. This implies that metaorders are not detected by other market participants, as assumed in some equilibrium models~\citep{farmer2013efficiency}. More importantly, the impacts of different metaorders are in fact \emph{dependent} and \emph{do not add up linearly}.
  \item To subtract the effect of the order flow, and to measure the marginal effect of \textit{one isolated} metaorder, we have selected metaorders that were not correlated to the market's local direction. This allowed us to show that for slow execution speeds the
  average execution cost ${\cal I}_{\rm exec}(Q,\mu)$ decreases -- even though measures are too noisy to make a proper fit --  and that the mechanical component of the permanent impact ${\cal I}^{\infty}_{\rm mec}(Q,\mu)$ is close to zero.
Such a behaviour is consistent with kernel models and statistical order book models that naturally generalize them.
\item The marginal impact of these isolated metaorders drops far below the $2/3$ level predicted by equilibrium theories. The permanent impact of isolated trades is probably zero, as concluded in \citet{brokmann2014slow} and \citet{gomes2015market}.
\end{enumerate}

\section{Conclusion}\label{sec:con}

Using a remarkable dataset which covers all transactions on the MtGox Bitcoin/USD exchange, we have conducted a comprehensive analysis of the market impact of over one million metaorders on Bitcoin. The Bitcoin market has two important features which motivate such a study: First, it corresponds to a single-asset economy, a quite unique example. Second, since each transaction is charged fees of $0.6\%$ of its amount (60 bps), the existence of market makers and arbitrageurs is the exception rather than the rule.

The fact that the square-root law for impact holds in these conditions allows us to assess some underlying hypotheses of current impact models. Due to the large fees, neither 
statistical arbitrage nor high-frequency market making can be at the origin of the square-root law (on the Bitcoin) and theoretical explanations of this stylized fact must be found elsewhere.
 Accordingly, the notion of fair pricing is irrelevant in this market, as it seems not reasonable to assume that investors behave in such a way that they break even on their impact (of order a fraction of a percent) when they have to pay fees that are much larger. 
Nor should one rely on any martingale conditions for the price, as the notion of ``price'' is not precisely defined under the scale of 60 bps due to the large spread. The fact that the square-root impact law holds on the Bitcoin market so precisely at all scales clearly demonstrates that equilibrium and arbitrage mechanisms are not the underlying mechanism of market impact. 
Furthermore, our study suggests that impact should be regarded as how trades dig into the opposite side rather than how they affect the ``price'' itself, which is not equivalent when the spread is not tight.


To conclude, this study incites us to think that impact is driven by a generic fundamental and local mechanism that emerges together with the simple phenomenon of supply and demand and pre-exist any notion of arbitrage (although it might remain compatible with it in the end as behaviours adapt to it). A promising avenue in our view, is the use of heterogeneous agent models \citep{donier2015walras}, in which supply and demand are the output of interactions in a complex system that cannot be reduced to a few representative agents, and which seem to reproduce impact pictures remarkably well \citep{DonierBonart14}.


\section*{Acknowledgements}

We warmly thank Jean-Philippe Bouchaud for countless discussions and for carefully reading the manuscript. We also thank N. Kornman and A. Tilloy for their precious insights on the Bitcoin and finally C.A. Lehalle and H.Waelbroeck for very useful discussions.

\bibliography{bibli}
\bibliographystyle{apalike}

\end{document}